\documentclass[12pt]{article}
\usepackage{amsfonts,amsmath,amssymb,epsf,cite}
\usepackage{graphicx,color,subfigure}
\usepackage[usenames,dvipsnames]{pstricks}
 \def\frac#1#2{{#1\over #2}}

 \def\ch{{\chi}}
\def\tch{{\tilde \chi}}

\def\cD{{\cal D}}
\def\cL{{\cal L}}

 \def\frac#1#2{{#1\over #2}}

 \def\ch{{\chi}}

\def\cP{{\cal P}}
 
 \def\cO{{\cal O}}

 \def\ben{\begin{equation}}
\def\een{\end{equation}}
\def\beq{\begin{eqnarray}}
\def\eeq{\end{eqnarray}}
\def\bea{\begin{eqnarray}}
\def\eea{\end{eqnarray}}
\def\vx{{\vec{x}}}

\begin{document}
\begin{titlepage}
\thispagestyle{empty}
\begin{flushright}
UK/11-06\\
\end{flushright}

\bigskip

\begin{center}
\noindent{\Large \textbf
{Quantum Quench across a Holographic Critical Point}}\\
\vspace{2cm} \noindent{
Pallab Basu\footnote{e-mail:pallab.basu@uky.edu} and
Sumit R. Das\footnote{e-mail:das@pa.uky.edu}}

\vspace{1cm}
  {\it Department of Physics and Astronomy, \\
 University of Kentucky, Lexington, KY 40506, USA}
\end{center}

\vspace{0.3cm}
\begin{abstract}

We study the problem of quantum quench across a critical point in a
strongly coupled field theory using AdS/CFT techniques. The model 
involves a probe neutral scalar field 
with mass-squared $m^2$ in the range $-9/4 < m^2 < -3/2$ in a
$AdS_4$ charged black brane background. For a given brane background
there is a critical mass-squared, $m_c^2$ such that for $m^2 < m_c^2$
the scalar field condenses. The theory is critical when $m^2 =
m_c^2$ and the source for the dual operator vanishes. At the critical
point, the radial operator for the bulk linearized problem has a zero
mode.  We study the dynamics of the order parameter with a time
dependent source $J(t)$, or a null-time dependent bulk mass $m(u)$
across the critical point. We show that in the critical region the
dynamics for an initially slow variation is dominated by the zero mode
: this leads to an effective description in terms of a Landau-Ginsburg
type dynamics with a {\em linear} time derivative. Starting with an
adiabatic initial condition in the ordered phase, we find that the
order parameter drops to zero at a time $t_\star$ which is later than
the time when $(m_c^2-m^2)$ or $J$ hits zero. In the critical region,
$t_\star$, and the departure of the order parameter from its adiabatic
value,
scale with the rate of change, with exponents determined by
static critical behavior. Numerical results for the order parameter
are consistent with these expectations.

\end{abstract}
\end{titlepage}
\newpage

\tableofcontents
\newpage

\section{Introduction and summary}
An important class of problems in quantum systems is that of quantum
quench, where a parameter in the hamiltonian varies with time,
typically attaining constant values at early and late
times \footnote{Sometimes quench is used to denote a sudden change. We
  will use this word to denote changes with arbitrary rates, in
  particular slow rates.}. Starting with some initial state, the
problem is to determine the nature of the final state. This problem
has recently attracted a lot of attention in several areas of
many-body physics, particularly because of progress in cold atom
experiments \cite{sengupta},\cite{CCa},\cite{CCc}.  Among other
things, this problem is interesting for two different reasons. The
first relates to the question of thermalization. Does the system
evolve into some kind of steady state ? If so, is the state "thermal"
in any sense ?  The second question deals with the situation where the
quench takes place across a value of the parameter where there is an
equilibriium critical point. In this case, Kibble-Zurek type scaling
arguments indicate that there are several physical quantities which
are {\em universal} and determined by the critical exponents of the
critical point. Furthermore for two dimensional theories which are
{\em suddenly} quenched to a critical point, powerful techniques of
boundary conformal field theory have been used in \cite{CCc} to show
that ratios of relaxation times of one point functions, as well as the
length/time scales associated with the behavior of two point functions
of different operators are given in terms of ratios of their conformal
dimensions at the critical point, and hence universal.  Another
related application of this phenomenon is in gravitational physics and
cosmology. In this context, this is the phenomenon of particle production in time dependent
backgrounds, which is relevant to many physical problems ranging from
black hole evaporation to behavior of quantum fluctuations in an
expanding universe.

There are very few theoretical tools available to study such systems
when they are strongly coupled. In this note we will explore the use
of AdS/CFT correspondence \cite{Maldacena} - \cite{AdSR} to this
problem.  AdS/CFT techniques have been in fact used in the past to
study quantum quench, though not at critical points.  In AdS/CFT the
boundary values of bulk fields become coupling constants, so that this problem
becomes that of determining a time dependent background with specified
initial and boundary conditions. As is usual, a {\em quantum} problem
in the boundary theory becomes a classical problem in General
Relativity.

One class of problems involve a vacuum initial state (in the bulk this
is pure $AdS$ in the distant past) and a change of the boundary value
of some field (e.g. the dilaton or the boundary metric) over some
finite time interval, in a regime where supergravity is always
valid. In the boundary field theory this corresponds to turning on a
time dependent source for the dual operator. Under suitable
conditions, this leads to black hole formation in the bulk
\cite{janik,otherthermalization,holoentanglement}.  The correlators at
future time would then be thermal with a temperature characterized by
the Hawking temperature. The time scale after which this happens
depends on the nature of the correlators, but turns out to be always
smaller than what one would expect from a conformally invariant system
evolving to a thermal state. Thus, in this case thermalization of the
field theory is signalled by black hole formation.  Another class of
problems involve a similar setup, but a suitable variation of the
coupling which prevents black hole formation in the supergravity
regime. The coupling, however, becomes {\em weak} at some time and the
bulk string frame curvature grows large, leading to a breakdown of the
supergravity approximation - thus mimicking a space-like singularity
\cite{Awad}. For the case of a slow variation of the coupling it turns
out that the gauge theory remains well defined and may be used to show
that a smooth passage through this region of small coupling is
possible without formation of a large black hole.  Related scenarios
appear in \cite{hertog} and \cite{eva}.

Many interesting phenomena in gauge-gravity duality, in particular
phase transitions, can be explored consistently in a {\em probe
  approximation}. In this approximation, a certain set of bulk fields
can be treated separately and independently of bulk gravity. One set
of examples are probe branes in a background $AdS \times S$. This
introduces hypermultiplets which in general live on a defect in the
original field theory. When the number of such branes, $N_f$ is much
smaller than the number of flux units which produce the background
geometry, $N_c$, the dynamics of these branes do not backreact on the
bulk metric which remains $AdS \times S$. The dynamics is then
described by a DBI action in a fixed background - the fields in this
action are then bulk probe fields. If the boundary value of such a
probe field is time dependent in a fashion similar to above, black
hole formation in the bulk is invisible in the probe
approximation. However, rather remarkably, thermalization of the
hypermultiplet sector is still visible. This manifests itself by the
formation of an apparent horizon (which evolve into an event horizon
in some situations) in the induced metric on the brane worldvolume
\cite{dnt, otherapparent}. Fluctuations of the brane feel this induced
metric and their correlators have thermal properties \footnote{For
related phenomena see \cite{Gursoy,wa,evans}}. 

None of the above works study quench dynamics near a phase
transition. In this paper we take the first step towards doing that,
in a simple model which displays an equilibrium critical point in a
probe limit \cite{liu1}.  The model involves a neutral scalar field
whose mass lies in the range $ -9/4 < m^2 < -3/2$ in the background of
a charged $AdS_4$ black brane. The overall coupling is chosen to be
large so that the scalar can be treated as a probe. For a given brane
background (i.e. for a given temperature and charge density in the
dual field theory), and for a vanishing non-normalizable mode of the
field (corresponding to a vanishing source of the dual operator) there
is always a critical mass $m_c^2$ below which the trivial solution is
unstable. In this regime the stable solution is nontrivial, signifying
a nonzero expectation for the dual operator in the dual theory. $m^2
=m_c^2$ with a vanishing source is then a critical point, which
happens to have standard mean field exponents.
This setup is in fact quite similar to models of
holographic superconductors and fermi surfaces
\cite{gubser}-\cite{fermisurface}.
However, here the scalar field is neutral
so that the dynamics of the gauge field is frozen. This model has been
argued to model phase transitions in antiferromagnets. As is well
known, in this mass range there are two inequivalent quantizations. Our
analysis is restricted to the conventional quantization of the bulk
scalar, where the mode which vanishes slower at the boundary is
treated as the source in the boundary theory. A similar analysis can
be easily performed for the alternative quantization.

We will study the time evolution of the order parameter starting from
adiabatic initial conditions in the ordered phase.  The passage
through the critical point is done in two ways - (i) by keeping the
bulk mass at its critical value and changing the leading term 
of the near-boundary expansion of the bulk field 
and (ii) by changing a bulk mass parameter with a vanishing
leading term.  The
field theory meaning of (i) is clear - this corresponds to turning on
a source dual to the bulk field : the critical point appears when the
source vanishes and the mass is at its critical value. The meaning of
(ii) is less direct - changing a bulk mass means changing the anomalous
dimension of the corresponding operator in the field theory. As
discussed in \cite{liu1}, one way to do this is in fact coupling this
scalar field to another field and changing the source for this other
field. 

Away from the critical point, adiabaticity holds for a sufficiently
low rate of change.  We will show that the adiabatic corrections are
proportional to {\em linear} time derivatives, even though the
equations of motion involve only second order time derivatives. This
is a well-known phenomenon and happens because of dissipation implied
by the presence of a horizon. In the bulk, we need to impose
regularity conditions at the horizon, which is easily done using
ingoing Eddington-Finkelstein coordinate \cite{bhmr,hubeny2} $u$ as
time. $u$ coincides with the usual time coordinate $t$ on the boundary
and $\partial /\partial u|_r = \partial / \partial t|_r$, though
$\partial / \partial_r|_u \neq \partial /\partial r|_t$. The bulk
equations now involve first order derivatives $\partial / \partial
u$ \footnote{Note that the analysis can be done using the usual time
  coordinate as well, as originally done in \cite{starinets}, but one
  has to take special care of the behavior of the solution near the
  horizon. The result is of course the same.}. As expected,
adiabaticity fails as one approaches the critical point at a time
which we choose to be $t=0$. The appearance of a first order time
derivative is related to the fact that the dynamical critical exponent
of similar models turn out to be $z=2$ \cite{holosupdynamical}.

When adiabaticity fails, the system enters a scaling regime.  A useful
way to understand the dynamics is to decompose the bulk field in terms
of eigenstates of the radial operator which appears in the linearized
problem around the equilibrium solution. For any nonzero temperature
of the background black brane, this operator has a zero mode which is
regular at the horizon exactly at the critical point.  We will show
that in the critical region, the dynamics for small $v$ is dominated
by that of this zero mode.  The other modes remain adiabatic. The
scaling properties can be then understood in terms of a
Landau-Ginsburg type dynamics of the zero mode with a {\em linear}
time derivative. The order parameter remains nonzero even when the
instantaneous value of the coupling reaches the equilibrium critical
value and first drops to zero at a time $t_\star \sim v^{-\alpha}$
where $v$ denotes the rate of change of the coupling in the critical
region. The departure of the value of the order parameter from the
instantaneous equilibrium value also scales as  $v^{\beta}$. The
exponents $\alpha$ and $\beta$ are essentially determined by the
static critical exponents.

Beyond the critical region, however, the other modes
become important. In fact, at very late times the bulk equation may be
approximated by its linearized form, so that the solutions are nothing
but the usual quasinormal modes around the background black hole. As
usual, the imaginary part of the lowest quasinormal mode frequency then
determine the decay of late time behavior.

We also present preliminary results of a numerical solution of the
time dependent equations for the bulk field and extract the time
dependence of the order parameter. The results are consistent with the
above picture. Our numerical results are not yet accurate enough to
verify the scaling behavior. However this should be possible with
further work.

In Section 2 we discuss spatially homogeneous classical quench in a
Landau-Ginsburg type model as a prelude. In Section 3 we review the
main faetures of the equilibrium phase transition. In Section 4 we
discuss the failure of adiabaticity in quantum quench across this
transition. We then show that in the critical region the dynamics for
an initially slow quench is dominated by that of the zero mode, 
and discuss its relationship of the dynamics to 
the Landau-Ginsburg model of Section 2. In Section 5 we
present the results of a numerical solution of the bulk equation of
motion and verify the expectations in the previous section. Section 6
contains some concluding remarks. The three appendices provide details
of derivation of the some of the results contained in the main text.

\section{Classical quench in a 0+1 dim model}

As a prelude to the main discussion, in this section we will study
classical spatially homogeneous quench in a Landau-Ginsburg model,
which is described by the equation,
\ben
\frac{d^{\alpha}\phi}{dt^{\alpha}}+m^2(t)\phi+\phi^{2 \beta-1}+J(t)=0
\label{22-2}
\een
As discussed above, holography converts 
the problem of a {\em quantum} quench to a problem
of {\em classical} quench in one more dimension - which is the
motivation behind studying (\ref{22-2}).
In a later section we will show that this
simple model describes some of the essential features of the dynamics
in the critical region of the holographic phase transition which is
the main subject of this paper. 
Note that we do not have any noise term (which is what is needed for
many other applications).

When $m^2$ and $J$ are time independent, this model
has a critical point at $m^2=J=0$.
Here we have kept  generic $\alpha$ and $\beta$ for 
convenience of later discussion. In most conservative models,
i.e. where energy is conserved, $\alpha=2$. However as we will see
$\alpha=1$ is more appropriate for the holographic transition which we
will discuss. In the following we will therefore use
this value. We will also concentrate on the $\beta=2$ case, i.e. a quartic interaction.

\subsection{$J$ Quench}

Let us first discuss the case where $J(t)$ is time varying, but a $m^2$
which is independent of time and positive.

For a  time independent $J$ we have, 
\ben m^2
\phi+\phi^3+J=0 
\een 
Denote the solution of this static equation
by $\phi_0(J,m)$. We then introduce a $J(t)$ which is slowly varying with
time. The idea is to start with adiabatic initial conditions at some
early enough time and study the time evolution of the order
parameter. To study the adiabatic expansion we write, 
\bea
\phi=\phi_0(J(t),m)+\epsilon \phi_1(t)+\cdots 
\eea
Here $\epsilon$ is an adiabaticity parameter introduced by performing
a scaling $t \rightarrow t/\epsilon$ so that any $\partial_t$ comes
with a factor of $\epsilon$. The ellipsis denote higher order terms in
$\epsilon$. 
To lowest order in $\epsilon$, the equation governing $\phi_1$ is,
\ben
\phi_1 (m^2+3 \phi_0^2(t))= \dot \phi_0(t) = \dot J(t) \frac{\partial \phi_0}{\partial J}
\een
The adiabatic expansion is good for a finite $m, \phi_0$ for a
sufficiently slowly varying $J(t)$.  The adiabatic expansion fails
when
\bea
& &\phi_1 \sim \phi_0 \\
&\Rightarrow& \frac{1}{m^2+3 \phi_0^2} \dot \phi_0 \sim \phi_0
\label{adbreak}
\eea
For a sufficiently slowly varying $J(t)$, this is
possible if both $m^2$ and $\phi_0$ are small, i.e. when we are close
to the critical point $m^2=J=0$.  Approaching the critical point along
the direction $m^2=0$ we have
$\phi_0(t)\approx (-J(t))^{\frac{1}{3}}$, and \ref{adbreak} leads to
the condition for breakdown of adiabaticity ,
\ben
\dot{J} \sim J^{\frac{5}{3}}.
\label{22-2a}
\een
If the quench is linear near the critical point, that is $J(t)\sim J_0
v t$ for small $J$, we get 
\bea vt^{\frac{5}{2}} \sim 1 
\eea 
as the condition for breakdown of adiabaticity.

When adiabaticity breaks down, the system enters a scaling region, In
the scaling region we use $J(t)=J_0vt+O (t^2)$. We rescale the field as: $
\phi \rightarrow \alpha \tilde \phi$ and $t \rightarrow \beta \tilde
t$. Choosing $\alpha^2\beta=1$ and $v\frac{\beta^2}{\alpha}=1$, we get
a $v$ independent equation for $\tilde\phi(\tilde t)$. The quadratic
or higher order temporal contribution in $J(t)$ only gives a
sub-leading contribution in $J_0 v$ in the scaling limit. Hence we have a
scaling solution, 
\ben 
\phi(t)=v^{\frac{1}{5}} \tilde
\phi(v^{\frac{2}{5}}t).  
\een 
This scaling solution leads to an estimate of the magnitude of
fluctuations ($\delta \phi(0)$) ,i.e. the departure of $\phi$ from the
equilibrium value at $t=0$.  Outside the critical region $\delta
\phi(0) \sim \frac{v}{m^2}$, whereas in the critical region $\delta
\phi(0) \sim v^{\frac{1}{5}}$. This also defines the ``zero crossing
time''($t_\star$) defined by $\phi(t_\star)=0$. We have $t_\star \sim
v^{-\frac{2}{5}}$ at the critical point. Which diverges in the
$v\rightarrow0$ limit. Out of criticality $t_{\star}$ actually
approaches a constant in the $v\rightarrow0$ limit.

\begin{figure}\label{LGsource1}
\begin{center}
\includegraphics[scale=0.75]{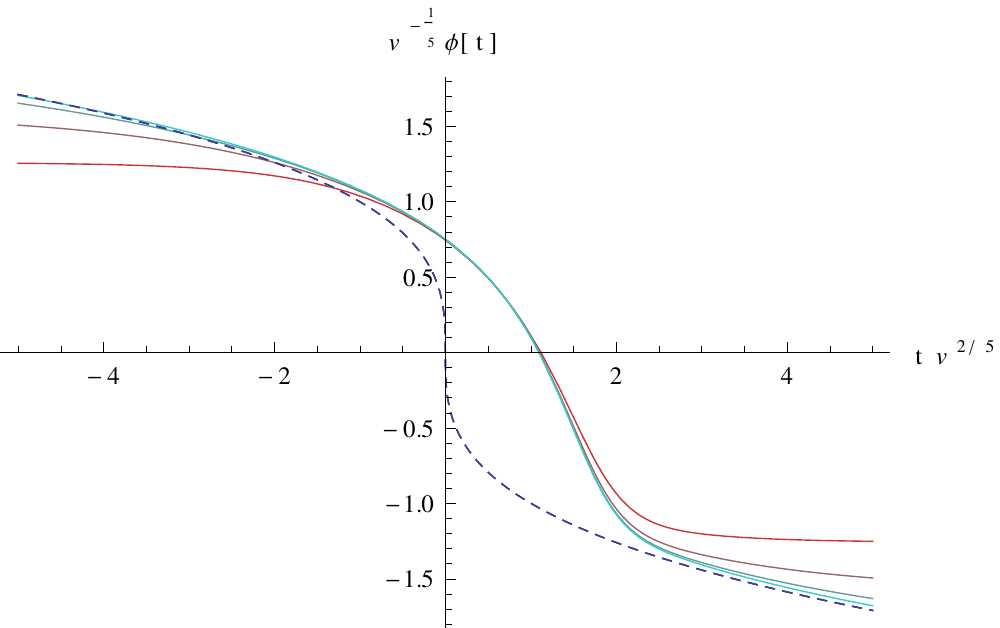}
\end{center}
 \caption{The order parameter as a function of time for a $J(t) =
   \tanh(vt)$ at $m^2=0$ with $v=10^{-0.5},10^{-1},10^{-1.5},10^{-2}$
   (from the bottom on the left). The adiabatic solution (dashed) is
   also shown as a comparison.}
\end{figure}

It is instructive to study the dynamics of the order parameter for a
profile of $J(t)$ which becomes a constant at early and late times and
behaves linearly with time in the critical region, for example 
\ben
J(t) = J_0 \tanh (vt) 
\een 
with $m^2=0$ for all times.  The numerical
solution of the equation of motion with an adiabatic initial condition
at early times is shown in Figure (\ref{LGsource1}) for a typical
value of $v$.

Numerically we see that at sufficiently early times the order
parameter tracks the adiabatic solution and overshoots the adaiabatic
solution as one approaches the critical region at $t=0$, falling down
to zero at some positive time $t_\star$. At later times, the solution
again approaches the adiabatic solution. The numerical solutions for
different values of $v$ are consistent with the behavior $\delta
\phi(0) \sim v^{\frac{1}{5}}$ and $t_\star \sim v^{-\frac{2}{5}}$ as
predicted by the scaling analysis.

For profiles of $J(t)$ which behave as some nontrivial power of $t$ in
the critical region, i.e. $J(t) \sim (vt)^n$ a similar analysis leads
to 
\ben \phi(t,v) = v^{\frac{n}{2n+3}} \tilde \phi(t
v^{\frac{2n}{2n+3}},1) \een leading to the scaling properties \ben
\delta \phi(0) \sim v^{\frac{n}{2n+3}}~~~~~~~~t_\star \sim
v^{-\frac{2n}{2n+3}}.  
\een

If there are more than one interaction terms, e.g. $\phi^{2\beta-1}$
present in the equation of motion, then the lowest order interaction
($\beta_0$) dominates the critical dynamics. In this case we have,
\ben 
\phi(t,v)=v^{\frac{n}{(2\beta_0-2)(n+1)+1}} \tilde \phi(t
v^\frac{(2 \beta_0-2) n}{(n+1)(2\beta_0-2)+1}).  
\een

\subsection{Mass quench}

In this section we will study classical quench across the critical point along the line $J=0$ by changing $m^2$ from negative to positive values.
For static $m^2<0$, the model has a non-trivial time-independent solution $\phi_0=\sqrt{-m^2}$. The static model has a second order phase transition near $m^2=0$. Now we introduce a time dependent mass function which  asymptotes to some constant value of $m^2 < 0$ at early times and then rise to a constant value $m^2 > 0$ at late times, e.g. 
\ben
m^2(t) =  -m_1^2 \tanh (vt)
\label{22-5}
\een
Once again we start with adiabatic initial conditions at some early enough time and study the time evolution of the order parameter. To study the adiabatic expansion we write,
\bea
\phi=\phi_0(t)+\epsilon \phi_1(t)+\cdots
\eea
where,
\bea
\phi_0(t)=\sqrt{-m^2(t)}
\label{0-1}
\eea
To lowest order in the adiabatic expansion
\ben
\phi_1 = \frac{\dot{\phi_0}}{2 m^2}
\een
Using the fact that $\phi_0 \sim \sqrt{-m^2)}$, we get that the adiabatic expansion breaks down as $\phi_1 \sim \phi_0$, i.e. when 
\ben
  \frac{\dot{\sqrt{(-m^2)}}}{(-m^3)} \sim 1.
\label{22-3}
\een
As expected this breaks down as we approach the equilibrium critical point. If  $(-m)^2 \approx v t$ near the critical point breakdown of adiabaticity happens when,
\ben
t \propto v^{-\frac{1}{2}}
\een

When adiabaticity breaks down the system enters a scaling regime. In
this regime, we can approximate $m^2(t) = - m_1^2 v t$. Here the
scaling relations are 
\ben 
\phi(t,v) \approx v^{\frac{1}{4}} \tilde
\phi(t v^{\frac{1}{2}},1)
\label{22-4}
\een
However, if the solution does not cross a phase transition
point then fluctuation around a static solution is suppressed
exponentially in $\frac{1}{v}$. It is also to be noted that if $m(t)$
is sufficiently slowly varying, the exact functional form of $m(t)$
is not needed. The only information we use to derive (\ref{22-4}) is that
$m(t)^2$ is a linear function of time near the phase transition and
$m(t) \rightarrow m_2$ as $t \rightarrow \infty$.

\section{The equilibrium phase transition in the holographic model}

The model of \cite{liu1} has a neutral scalar field $\phi(t,r,\vx)$ in
the background of a charged $AdS_4$ black brane. The lagrangian is
given by 
\ben 
\cL =
\frac{1}{2\kappa^2\lambda}\sqrt{-g}[-\frac{1}{2}(\partial \phi)^2
  -\frac{1}{4}(\phi^2+m^2)^2-\frac{m^4}{4}]
\label{1-1}
\een
The background metric is given by (in $R_{AdS}=1$ units)
\ben
ds^2 = [-r^2 f(r)dt^2+r^2 d\vx^2]+\frac{dr^2}{r^2 f(r)}
\label{1-2}
\een
where
\ben
f(r) = [1+\frac{3\eta r_0^4}{r^4}-\frac{1+3\eta r_0^3}{r^3}]~~~~~~~~0 \leq \eta \leq 1
\label{1-3}
\een
The associated Hawking temperature is then given by
\ben
T = \frac{3}{4\pi r_0}(1-\eta)
\label{1-4}
\een
In the following we will replace $r \rightarrow r r_0$.

One should add the Einstein-Hilbert action and a cosmological constant
term to (\ref{1-1}). In the limit of large $\lambda$, however, the
field $\phi$ can be regarded as a probe field since its back reaction to
the metric via Einstein's equations is small. This is the
approximation we adopt. 

In \cite{liu1} it was shown that when the mass lies in the range 
\ben
-\frac{9}{4} < m^2 < -\frac{3}{2}
\label{1-5} 
\een 
there is a critical phase transition at
some value of $T = T_c(m)$ when the source to the dual operator
vanishes.  
Conversely, for a given $T$ there is a
value of $m^2 = m_c^2$ where the theory is critical.

The upper limit in (\ref{1-5}) is the BF
bound for the near-horizon $AdS_2$ geometry which appears in the
extremal ($\eta = 0$) metric. (Note that the AdS scale for this
infrared $AdS_2$ is given by $1/\sqrt{6}$ in our units). The lower
bound is the BF bound for the asymptotic $AdS_4$.  Field
configurations which are translationally invariant in the $\vx$
directions satisfy the equations of motion 
\ben
\frac{1}{r^2}[-\frac{r^2}{f(r)}\partial_t^2+\partial_r(r^2
  f(r)\partial_r)]\phi -m^2 \phi -\phi^3 = 0 
\label{1-6}
\een 
Near the $AdS_4$
boundary the asymptotic behavior of the solution to the linearized
equation is of the form \footnote{When we turn on $J(t)$, it is a valid concern whether we will be able to neglect the non-linear term near the boundary. This can be done as long as $\Delta>0$ or $m^2<0$.}
\ben
\phi (r) = J(t)r^{-\Delta_-}[1+O(1/r^2)]+<{\cal O}>(t) r^{-\Delta_+}[1+O(1/ r^2)]
\label{1-7}
\een
where $\Delta$ is given by
\ben
\Delta_\pm = \frac{3}{2} \pm \sqrt{m^2 + \frac{9}{4}}
\label{1-8}
\een
In the range of masses of interest, both the solutions are
normalizable, so that there is a choice of quantization. The standard
quantization considers the coefficient $J(t)$ is the source in the
dual field theory and $<{\cal O}>(t)$ then gives the expectation value of the
dual operator. In the alternative quantization the expectation and source change the role.

Consider first the linearized problem, ignoring the cubic term. By a standard change of coordinates to tortoise coordinates $\rho$ and a field refinition to $\chi$,
\ben
d\rho =- \frac{dr}{r^2 f(r)}~~~~~~~~~~~~~~\phi (r,t) = \frac{\chi (\rho,t)}{r}    
\label{1-9}
\een
The horizon is then at $\rho = \infty$ and the boundary is at $\rho = 0$. 
the equation becomes (\cite{Basu2})
\ben
-\partial_t^2 \chi = -\partial_\rho^2 \chi + V_0(\rho) \chi \equiv \cP_\rho \chi
\label{1-10}
\een
with
\ben
V_0(\rho) = f(r)[(m^2+2) - \frac{6\eta}{r^4}+\frac{1+3\eta}{r^3}]
\label{1-11}
\een
where in $V_0(\rho)$ we need to express $r$ in terms of $\rho$ using
(\ref{1-9}).  

For solutions of the type $\chi \sim e^{-i\omega t}$,
equation (\ref{1-11}) is a Schrodinger problem in a potential
$V(\rho)$. The potential goes to zero at the horizon $\rho = -\infty$
and behaves as $\frac{(m^2+2)}{\rho^2}$ near the boundary $\rho =
0$.  Note that for a brane background at any finite temperature, $f(r) \sim
(r-1)$ near the horizon, while $\rho \sim -\log (r-1)$ so that $V_0
\sim e^{-\rho}$ as we approach the horizon. In contrast, for the
extremal background $f(r) \sim (r-1)^2$ while $\rho \sim 1/(r-1)$ so
that $V_0 \sim 1/\rho^2$. This makes the analysis for the extremal
background rather subtle. In this paper we will work with the
non-extremal case.

Due to the vanishing of the potential near the horizon the
time-independent problem has a continuum of modes starting at
$\omega=0$. The behavior of the solution near the horizon $\rho
=-\infty$ is of the form $e^{-i\omega(t\pm\rho)}$, the two signs
representing ingoing and outgoing waves.  We are eventually interested
in solving the problem with {\em ingoing} boundary conditions at the
horizon.

Generically for any value of $m^2$ there are quasi-normal modes with
complex frequencies. These are normalizable modes with ingoing
boundary condition. Unlike the continuum discussed above these modes
form a set of disconnected poles in the complex frequency plane.
However, normalizable bound states appear when $m^2$ is sufficiently
negative : this is what has been shown in \cite{liu1}. 

\begin{figure}
\begin{center}
 \includegraphics[scale=0.65]{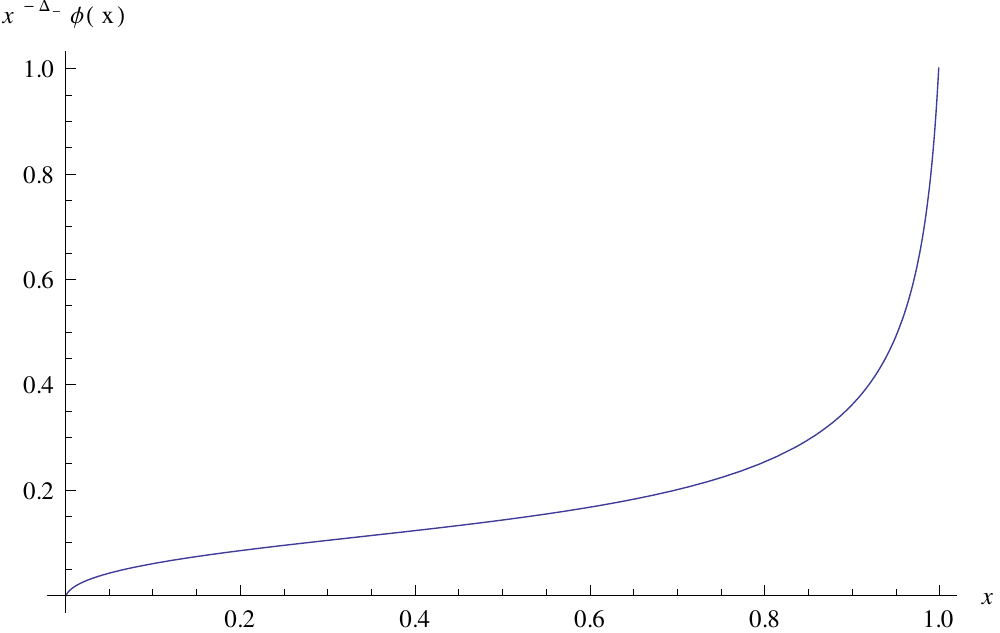}
\end{center}
\caption{Profile of (rescaled) zero mode ($x^{-\Delta_-} \phi(x)$, where $x\sim
  \frac{1}{r})$) with $m^2 \approx -2.235 $ and $\eta=0.98$.}
\label{zmd}
\end{figure}

There is a critical value of $m^2 = m_c^2$ when a zero energy bound
state appears, which vanishes in an appropriate fashion at the
boundary and is in addition purely ingoing at the horizon.  In the
complex frequency plane some quasinormal mode(s) hit the origin at
$m^2 = m_c^2$. This critical mass is $m_c^2 = -\frac{3}{2}$ when $\eta
= 1$ and decreases with decreasing $\eta$ or increasing
temperature. The existence of such a bound state for $m^2=-2$ has been
shown by variational methods in \cite{hhh}.  A similar demosntration
should be possible for other values of $m^2$.  We have numerically
verfied the existence of this zero mode. A typical plot the zero mode
is shown in in Fig \ref{zmd}. Note that the leading large-r behavior has been
factored out to ensure that the source term indeed vanishes at the boundary.

The presence of a bound state in the Schrodinger problem means that
the solution $\chi =0 $ of the full nonlinear problem is unstable, and
one has to look for other nontrivial solutions. In \cite{liu1} it was
shown that such a stable nontrivial static solution $\phi_0 (r)$
exists in the range of masses given by (\ref{1-5}), both for the
standard and alternative quantizations. This means that the
expectation value of the dual operator is nonzero, i.e. the field
condenses, and $m^2 = m_c^2(T)$ is a critical point . One can approach
this critical point either by fixing the mass and tuning the
temperature, or by fixing the temperature and tuning the mass. The
latter may appear to be somewhat strange from the point of view of the
boundary field theory. However, it is only the IR mass which is
relevant - and one may imagine obtaining this by coupling $\phi$ to
another field $\Psi$ by $\Psi \phi^2$. The expectation value of $\Psi$
which may results from changing a coupling in the boundary field
theory then renders a mass to $\phi$ \cite{liu1}.

The critical behavior has standard mean field exponents at any finite
$T$. If the operator dual to the field $\phi$ is $\cO$ and the source
is $J$ then an analysis identical to that presented in \cite{liu1}
leads to (for $m^2-m_c^2 \rightarrow 0^+$) \ben <\cO>_{J=0}\sim
(m^2-m_c^2)^{1/2}~~~~~\frac{d<\cO>}{dJ}|_{J=0}\sim
(m^2-m_c^2)^{-1}~~~~~~<\cO>_{m=m_c}\sim J^{1/3}
\label{1-12}
\een 
Exactly at zero temperature the phase transition is of BKT type
and the order parameter depends exponentially 
\ben <\cO>_{J=0} \sim
\exp \left[-\frac{\pi\sqrt{6}}{2\sqrt{m_c^2-m^2}}\right]
\label{1-13}
\een

The zero mode of the linearized operator will play a key role in the
following. This zero mode is known to exist for any non-zero
temperature. At zero temperature the situation is less clear. For a
similar zero temperature system involving the D3-D5 system \cite{bktpapers}
such a zero mode does not exist \cite{jensen}.

\section{Quench across the critical point the breakdown of adiabaticity.}

Our aim is to study quench dynamics across this equlibrium phase
transition. In a quench situation, some parameter of the boundary theory
is made time dependent. The first issue we will study is the breakdown
of adiabaticity as we approach the critical point.  First, we argue
that away from the critical point, the order parameter tracks the
changing coupling adiabatically.  We will then determine the regime near the critical
point where adiabataicity breaks and show that in this region the
order parameter and the time scale are proportional to some power of $v$ pretty
much like the model studied in section 2.

\subsection{Quenching the source}

As discussed above, a simple way to quench across the critical point
is to remain at the critical value of the mass and consider a time dependent source for the dual operator in the
boundary field theory. This corresponds to a nontrivial boundary value
for the bulk field. Our discussion is in the conventional quantization
of the theory, i.e $J(t)$ in (\ref{1-7}) is treated as a source. The
case of alternative quantization should follow along identical lines.

It is well known that to study low frequency modes in the background
of a black brane it is convenient to use ingoing Edddington-Finkelstein
coordinates,  
\ben
u=\rho - t,~~~~~~\rho
\een
where $\rho$ is defined in (\ref{1-9}).
In terms of these coordinates the equation of motion \ref{1-6}) becomes
\ben
-2\partial_u\partial_\rho \chi = -\partial_\rho^2 \chi + V(\rho,\chi).
\label{4-1}
\een
where
\ben
V(\rho,\chi)= V_0(\rho)\chi+\frac{f(r)}{r^2}  \chi^3.
\een

This equation has to be solved with the boundary condition that the
field is {\em regular at the horizon}, which at the linearized level
is equivalent to requiring that the waves are purely ingoing at the
horizon \cite{bhmr,hubeny2}. 
At the linear level the analysis can be of course performed in $(t,\rho)$
coordinates as well. However, in this case one needs special care to extract a
leading singular piece before performing the low energy expansion
\cite{starinets}. At the full nonlinear level this procedure is possibly rather involved, but imposing regularity in EF coordinates remain simple.

We need to solve (\ref{4-1}) with the condition
\ben
\chi(u,\rho) \rightarrow \rho^{-1+\Delta_-} J(u) ~~~~~~{\rm as}~~ \rho
\rightarrow 0
\label{4-1-1}
\een
where $\Delta_\pm$ are defined in (\ref{1-8}).
We have considered the source to be a function of $u$ rather than a
function of $t$. On the boundary $\rho =0$ $\rho$ and $t$ are the same
(and quite generally $\partial / \partial u |_\rho = \partial /
\partial t |_\rho$), so that this does correspond to a time dependent
source in the boundary theory. However, as we will soon see, it is
convenient to use (\ref{4-1-1}). This form makes bulk causality
explicit. 

To perform the adiabatic expansion, let us decompose the field $\chi
(\rho, u)$ as
\begin{align}
 \chi(\rho,u)=\chi_l(\rho,u)+\chi_s(\rho,u) 
\end{align}
Where $\chi_l(\rho,u)=J(u) \rho^{-1+\Delta_-}$ and $\chi_s(\rho,u)
\sim \rho^{-1+\Delta_+}$ as $\rho \rightarrow 0$.  We get,
\begin{align}
[-\partial_\rho^2+ V_0(\rho)] \chi_s + \frac{f(r)}{r^2}( \chi_s^3+3
\chi_l \chi_s^2+3 \chi_l^2 \chi_s)=- [-\partial_\rho^2 + V_0(\rho)]
\chi_l-\frac{f(r)}{r^2}\chi_l^3 \\ \nonumber - 2
\partial_u\partial_\rho \chi_l - 2 \partial_u\partial_\rho \chi_s
\end{align}
If $\chi_l(\rho,u)=\chi_l(\rho)$ is time independent we have a static
solution $\chi_s(\rho,u)=\chi_0(\rho)$. This has been discussed in the
previous section. When we turn on a time dependent $J(u)$ which is
{\em slowly varying}, one expects that the time dependent solution of
(\ref{1-6}) may be constructed in an adiabatic expansion. To
understand that we write, 
\ben \chi_s(\rho,u) = \chi_0(\rho, J(u)) +
\epsilon ~\chi_1(\rho,u) + \cdot.  
\een 
Here $\epsilon$ is an adiabaticity parameter which keeps track of the adiabatic
expansion. If we scale $u \rightarrow u/\epsilon$, each $u$ derivative
is of order $O(\epsilon)$. The ellipsis in (\ref{1-15}) are higher
order terms in $\epsilon$. The idea then is to insert (\ref{1-15})
into the equations of motion and obtain equations for $\chi_1,
\chi_2,\cdots$ order by order in $\epsilon$. To the lowest order one
gets 
\ben
\cD_\rho^{(1)} \chi_1 =   \{  [-\partial_\rho^2+ V_0(\rho)]  +\frac{f(r)}{r^2}
    (3\chi^2_0+6 \chi_l\chi_0+3\chi_l^2) \} \chi_1=+2
    \partial_u\partial_\rho \chi_l +2 \partial_u\partial_\rho \chi_0
\label{1-16} 
   \een
In deriving (\ref{1-16}) we have ignored any time dependence of
$\chi_1$, as is appropriate in an adiabatic expansion.  Generically, at
each order in perturbation we may cast the adiabatic problem as,
\ben
\cD_\rho^{(n)} \chi_n={\cal J}_n(u,\rho),
\label{1-17}
\een
where ${\cal J}_n(u)$ is a source term contains time derivatives and
determined by the lower order equations in an adiabatic expansion. 

As
discussed before,  near the black hole horizon the differential
operator becomes a plane wave operator and the potential
part vanishes. Since the potential $V_0(\rho)$ goes to zero at the horizon $\rho = \infty$ and the terms which involve $\chi_0, \chi_l$ in (\ref{1-16}) also vanish at the horizon (because of the overall factor of $f(r)$), the operator $\cD_\rho$ has a continuous spectrum which begins at zero.  It is, therefore, a non-trivial fact that there is adiabaticity away from the critical point. We will now show why this is so and how adiabaticity breaks down near the critical point.

 These equations are most conveniently solved by Green's
function method, 
\ben \chi_n=\int d\rho' G^{(n)}(\rho,\rho') {\cal
  J}_n(u,\rho').
\label{1-17a}
\een
where $G^{(n)}(\rho,\rho^\prime)$ is the Green's function of the operator
$\cD_\rho^{(n)}$ with the boundary conditions $G(0,\rho) =0$ and
$G(\infty,\rho)$ is regular.

Let us concentrate on the lowest order terms in the adiabatic
expansion, $n=1$. All higher terms may be found out by similar procedure.
The Green's function is given by,
\begin{align}
G^{(1)}(\rho,\rho') = \frac{1}{W(\chi_1,\chi_2)} \, \chi_1(\rho')
\chi_2(\rho) , \quad \rho < \rho' \\
	     =  \frac{1}{W(\chi_1,\chi_2)} \, \chi_2(\rho')
             \chi_1(\rho), \quad \rho > \rho',
\end{align}
where $\chi_1$ and $\chi_2$ are solutions of homogeneous part of eqn
(\ref{1-17}) satisfying appropriate boundary condition at the horizon
$\rho =\infty$ 
and the boundary $\rho = 0$ 
respectively, and $W(\chi_1,\chi_2)$ is the Wronskian which is
independent of $\rho$ in this case. We have
normalized $\chi_1(\rho)$ and $\chi_2(\rho)$ in such a fashion that
$\chi_1=1$ at the horizon and $\chi_2 \rightarrow \rho^{-1+\Delta_-}$
near the boundary. 

Given that we are looking for a solution which is regular at the
horizon ($\rho \rightarrow \infty$), $\chi_1$ is a constant at the
horizon and so is the Green's function.  Nevertheless, the integral in
eqn (\ref{1-17a}) is is finite. This is because regularity of the
functions $\chi_l$ and $\chi_0$ mean that $\partial_r \chi_l,
\partial_r \chi_0$ are finite at the horizon.  Since $(r-1) \sim
e^{-\rho}$, this implies that ${\cal J}(u,\rho) \sim
\partial_\rho(\chi_l+\chi_0) \sim \exp(-\rho)$, damping out any
divergence in (\ref{1-17a}) from the large $\rho$ region.

In general, near the
horizon  $\chi_2$
can be expressed as a linear combination of a regular
and irregular solution, i.e $\chi_2(\rho \rightarrow \infty)=a \rho +
b$. It can be easily checked that $W(\chi_1,\chi_2) =a$. 
Hence, 
\bea \chi_1(\rho,u)=
-\frac{1}{a} \int d \rho' \left[ \theta(\rho-\rho') \chi_1(\rho')
\chi_2(\rho)+ (\rho \rightarrow \rho') \right]  (2 \partial_u \partial_\rho
\chi_0+2 \partial_u\partial_\rho \chi_l) 
\eea
This is finite as long as $a$ is finite, and small for sufficiently
small $\partial_u \chi_0$ and $\partial_u \chi_l$.

At the phase transition point, however, the operator $\cD_\rho^{(1)}$
has a zero mode which is regular at the horizon. This is because at
$J=0$ the factors involving $\chi_0$ and $\chi_l$ in the left hand
side of (\ref{1-16}) vanish, and the operator is identical to the
operator acting on the linearized small fluctuations at $m^2 = m_c^2$
around the trivial solution $\chi_0 = 0$, i.e. the operator $\cP_\rho$
which
appears on the right hand side of (\ref{1-10}). We know that this
operator has a zero mode which is regular at the horizon {\em and}
vanishes as $\rho^{-1+\Delta_-}$ at the boundary $\rho = 0$. This
means that at this point $a = 0$. Therefore, the first adiabatic
correction diverges.

To derive the precise condition for adiabaticity, it is sufficient to
look at the operator $\cD_\rho^{(1)}$ for small $J(u)$.  For small
$J(u)$, the leading departure from the critical operator comes from
the term which is proportional to $\chi_0^2 \sim J^{2/3}$.  Thus we
can use perturbation theory in $J$ to estimate $a \propto J^{2/3}$. As
argued before, $\chi_0 \sim (-J)^\frac{1}{3}$, while $\chi_l \sim J$.
Hence the leading divergence in $\chi_1$ can be estimated as 
\ben
\chi_1(\rho,u) \sim \frac{J^{-2/3}\dot{J}}{J^{2/3}} = J^{-4/3}\dot{J}.
\label{1-20a}
\een
The breakdown of adiabaticity happens when,
\bea
\chi_1(\rho,u) \sim \chi_0 \\
\Rightarrow \dot{J} \sim J^{5/3},
\label{1-20b}
\eea
exactly as in the $0+1$ dimensional model (equation (\ref{22-2a}).
Assuming a linear quench, i.e $J(u) \approx v u$, we get 
\bea 
vu^{\frac{5}{2}} \sim 1 
\eea 
as the condition for breakdown of adiabaticity. 

\subsection{Scaling near the phase transition}

In this section we argue that the critical region dynamics is dominated by that of the zero mode for (initially) small quench rate. We will demonstrate this for the case where $J(u) = vu$ in the critical region. Extension to non-linear quenches is straightforward.

To understand this let us rescale $\chi \rightarrow v^\frac{1}{5} \tilde \chi_s,u \rightarrow v^{-\frac{2}{5}}\tilde u.$. Assuming a linear quench, i.e. $\chi_l \sim v u \tilde \chi_l(\rho)$, we have at the leading order in $v$,
\begin{align}
 [-\partial_\rho^2+ V_0(\rho)] \tch_s+v^{\frac{2}{5}} [\frac{f(r)}{r^2} (\tch_s)^3 + \tilde u [-\partial_\rho^2+ V_0(\rho)] \tch_l+2 \partial_{\tilde u}\partial_{\rho} \tch_s ] +\cdots=0
\label{veq}
\end{align}
The ellipsis denote terms which contains higher powers of $v$.

Now let us expand the sub-leading part of the scalar field in eigenfunctions of the operator $\cP_\rho$ (defined in equation (\ref{1-10}) at  the critical point,
\begin{align}
\tch_s(\rho,u) = \int \tilde a_k(u) \ch_k(\rho)  dk 
\end{align}
where $\ch_k$ satisfy
\ben
\cP_\rho^c \chi_k = [-\partial_\rho^2+ V_0^c(\rho)]\chi_k = k^2 \chi_k
\een
where $V_0^c$ denotes the potential in (\ref{1-11}) at $m^2=m_c^2$.  the eigenfunctions $\chi_k(\rho)$ are delta function normalized. The eigenfunctions obey the condition 
\ben
{\rm Lim}_{\rho \rightarrow 0} [ \rho^{-\Delta_-} \chi_k (\rho) ] = 0
\een

In terms of the eigen-coefficients $a_k(u)$ the equation (\ref{veq}) becomes, 
\begin{align}
 \nonumber & k^2 \tilde a_k+ v^{\frac{2}{5}}\left( \tilde u  {\cal J}_{k} -\int b_{kk'} \partial_{\tilde u} \tilde a_{k'} dk'- \int \tilde a_{k'} \tilde a_{k''} \tilde a_{k'''} C_{k,k',k'',k'''} d k' dk'' dk''' \right)=0
\label{modeeqn}
\end{align}
where,
\begin{align}
\nonumber  {\cal J}_{k}& =\int \ch_k (\rho) [-\partial_\rho^2+ V_0(\rho)] \ch_l \,  d\rho \\
b_{kk'}=\int d \rho \, \ch_k \partial_\rho \ch_k' &\,\,\, , \, C_{k,k',k'',k'''}=\int d\rho \, \ch_k \ch_{k'}  \ch_{k''} \ch_{k'''} \frac{f(r)}{r^2} .
\end{align}

The equation (\ref{modeeqn}) suggests that there is a solution in a perturbation expansion of powers of $v^\frac{2}{5}$. Writing
\begin{align}
 \nonumber \tilde a_k(\tilde u)&=\delta(k) \tilde \xi_0(\tilde u) + v^{\frac{2}{5}} \tilde \eta_k(\tilde u)+\cdots,
\end{align}
we get an equation for $\tilde \xi_0$,
\begin{align}
\tilde u \, {\cal J}_{0} - b_{00} \frac{d}{d\tilde u} \tilde \xi_0(\tilde u)-C_{0000} \tilde \xi_0(\tilde u)^3=0
\label{zmeqn}
\end{align}
and a set of equations for $\tilde \eta_k$
\begin{align}
 \tilde \eta_k(\tilde u) &=-\frac{1}{k^2}\left (\tilde u {\cal J}_{k}- b_{k0} \frac{d}{d\tilde u} \tilde \xi_0(\tilde u)-C_{k000} \tilde \xi_0(\tilde u)^3\right) \\
 &= -\frac{1}{k^2}\left ( \tilde u ({\cal J}_{k}-{\cal J}_{0}) -
 (b_{k0}-b_{00} )\frac{d}{d\tilde u} \tilde \xi_0(\tilde
 u)-(C_{k000}-C_{0000}) \tilde \xi_0(\tilde u)^3\right)
\label{scalingeqn})
\end{align}
In the last line of the above equation we have subtracted the equation for $\chi_0(u)$ to make the $k \rightarrow 0$ limit explicit. 

If the values of $k$ were discrete, the $k=0$ mode would clearly dominate the dynamics for small $v$. However, $k$ is a continuous parameter starting from zero - this naively indicates that the perturbation expansion in $v^\frac{2}{5}$ could break down. Whether this happens or not depends on the behavior of the numerator on the right hand side of (\ref{scalingeqn}).

It turns out, however, that  all the terms like $({\cal J}_{k}-{\cal J}_{0}) \sim k^2$ as $k\rightarrow 0$, 
rendering the perturbation expandion in $v^{\frac{2}{5}}$ well defined.
This is shown in Appendix (\ref{kscaling}). 

Thus, in the critical region the solution $\chi_s$ is dominated by the zero mode piece proportional to $\tilde \chi_0$. In terms of the original variables, we have finally
\begin{align}
 \chi_s(\rho,u) \approx v^{\frac{1}{5}}\tilde \xi(v^{\frac{2}{5}}u) \chi_0(\rho)+v^{\frac{3}{5}} \int \tilde \eta_k(v^{\frac{2}{5}}u) \chi_k(\rho) dk
\end{align}
Since the equation for $\tilde \xi_0$ is identical to the equation for the quantity $\phi$ in the toy model (\ref{22-2}), we are led
to scaling relations identical to the $0+1$ dimensional model. The one point function of the dual operator is given by 
\ben
<\cO>(u)  \sim {\rm Lim}_{\rho \rightarrow 0} [\rho^{1-\Delta_+} \chi_s (\rho,u)]
\een
Therefore, the time dependence of $\cO$ is governed by the same scaling behavior.

The above analysis can be carried out for $m^2$ away from the critical value as well. In this case, the terms like ${\cal J}_{k} \sim k$ as $k\rightarrow 0$, so that the expansion in $v^{5/2}$ is not valid. This means that away from the critical point, all the modes are important.

\subsection{Mass quench}

Another way to quench across the critical point is to consider a time dependent $m^2$ with no source, keeping the background geometry
fixed. The mass function is taken to asymptote to some constant value
of $m^2 < m_c^2$ at early times and then rise to a constant value $m^2
> m_c^2$ at late times, e.g.  
\ben 
m^2(u) = m_c^2+(m_c^2-m_1^2)\tanh (v u)
\label{2-1}
\een
Note that we have a mass which depends on the EF coordinate $u$ rather
than the time coordinate $t$. The motivation for this is as
follows. As mentioned earlier, we may consider the bulk mass term as
arising from a coupling of the scalar field with some other
field. This latter field acquires an expectation value in response to
a source which leads to a mass of our field $\phi$. In this scenario,
a time dependent source of the second field would lead, by causality,
to an effective mass-squared which has the form $m^2(u) G(\rho) +
\cdots $, where the ellipsis denote terms which involve $u$
derivatives of $m(u)$. The function $G(\rho)$ vanishes at the boundary
and is regular at the horizon. To leading order we therefore get an
equation like (\ref{4-1}) with the potential $V_0$ modified by
replacing $m^2 \rightarrow m^2(u) G(\rho)$.

The analysis in the previous sections can be now easily repeated for
this problem. Below we outline the main steps and results.

Let us consider the case of a time dependent mass which asymptotes to a
sufficiently negative value, $m_1^2$. If the mass remained constant at
$m_1^2$ there is a nonzero static solution
$\chi_0(r,m_1)$. Now once we make the the mass time dependent we solve
(\ref{4-1}) adiabatically in $\partial_u$. For a mass which is
    {\em slowly varying} one expects that the time dependent solution
    of (\ref{1-6}) may be constructed around an adiabatic solution,
\ben 
\chi(\rho,u;m(u)) = \chi_0(\rho,
    m(u)) + \epsilon^2~\chi_1(\rho,u) + \cdot.
\label{1-15}
\een 
$\chi_0(\rho;m)$ is the static solution obtained by ignoring the time
dependence of the mass completely.  Here $\epsilon$ is an adiabaticity
parameter which keeps track of the adiabatic expansion.

Following similar arguments like $J$ quench case , the divergence in
$\chi_1$ can be estimated to yield
\ben
\chi_1(\rho,u) \sim \frac{1}{\delta m^2} \partial_u (\sqrt{\delta m^2(u)})  
\label{1-20c}
\een
where $\delta m^2 \equiv (m_c^2 - m^2)$. The condition for break down
of adiabaticity for linear quench ($ \delta m^2 (u) \sim vu$ near the
critical point) is similar to $0+1$ dim case, 
\ben t v^{\frac{1}{2}}
\sim 1 
\een

When adiabiticity breaks down, the system enters a scaling
regime just like the $0+1$ dim case.  Following arguments similar to
what discussed in the previous section we find,
\begin{align}
 \chi(\rho,u) \approx v^{\frac{1}{4}}\tilde \xi_0(v^{\frac{1}{2}}u)
 \chi_0(\rho)+v^{\frac{3}{4}} \int \tilde b_k(v^{\frac{1}{2}}u)
 \chi_k(\rho) dk
\end{align}
The first term defines the scaling relation. The expectation value of
the scalar field in the scaling regime has the following properties,
\ben
<\cO(t)>=<\cO(t v^\frac{1}{2})> v^{\frac{1}{4}}
\een

\section{Numerics}

\begin{figure}
\centering \subfigure[The plot is for $\phi(x,t)$, ($x\sim
  \frac{1}{r}$). It is to be noted how the disturbance propagates
  following a light
  cone.]{\includegraphics[scale=0.6]{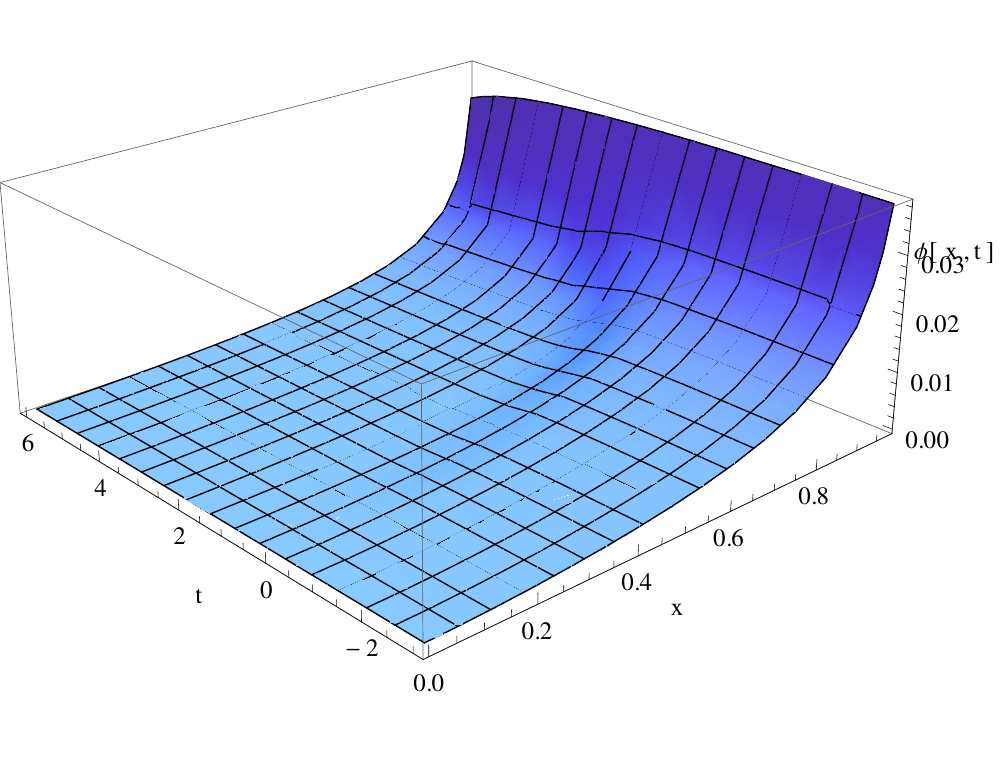}}
\hspace{1cm} \subfigure[Plot of boundary expectation value $<{\cal
    O}>(t)$.]
{\includegraphics[scale=0.5]{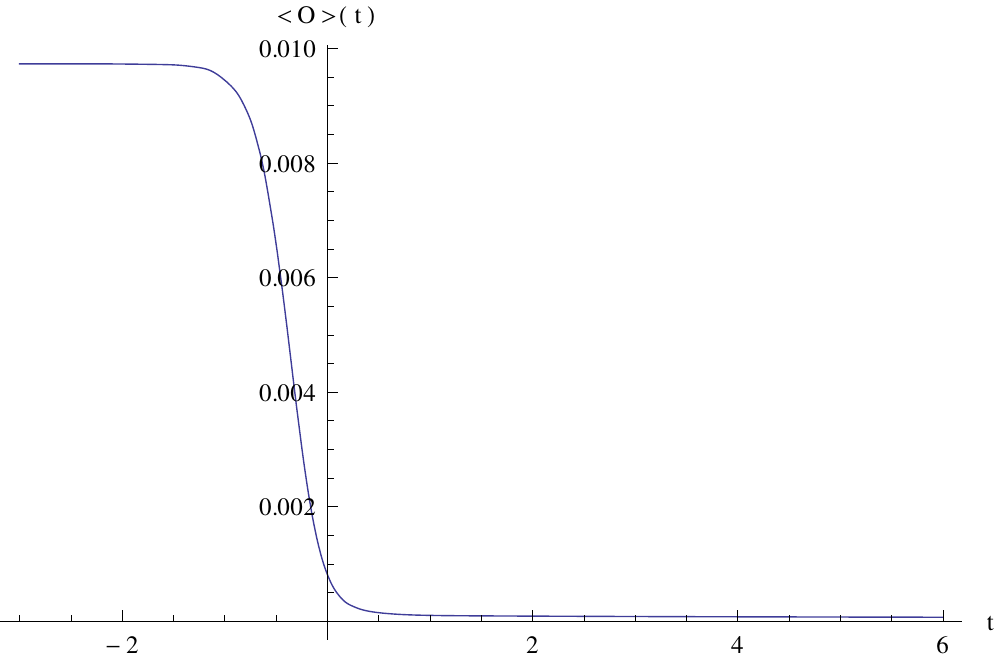}}
\caption{A typical response under a continuous but fast change of
  $m^2(t) \approx -2.1825 (0.9-0.1\tanh (3 t)) $. Here $\eta \approx
  0.991$ and $m_c^2 \approx -2.169$ at $t=0$.}
\label{dasbasu}
\end{figure}

In this section we will solve the equations (\ref{1-10}) numerically,
although our analysis is quite preliminary. Mainly we would like to
understand how the system behaves away from the regime of slowly
varying quench. It is generally hard to solve PDE's
numerically. However we will use the convenient method of
lines. In this method the PDE is at first discretized only in the $r$
direction. This gives rise to a set of ODEs. This set of ODEs is then
treated in a ODE solver. We fix the boundary condition at the two
ends, at the boundary and at the horizon. We use a regularity boundary
condition at the horizon and appropriate standard quantization
condition at infinity. The initial radial profile of the scalar field
and its derivative are given at $t=0$. In most cases the initial value
scalar profile would be a solution with $m^2=m^2(t_0)$. Hence
$m^2<m^2_c$. Then we will change $m^2$ to some other value higher than
the critical mass. The first point to note is that at late times , i.e. out of
the critical region, any finite energy excitation is gradually sucked up
by the black hole.  In fact at late times, the equation can be approximately
linearized and the solution 
in a black hole background can be expressed as a sum of quasi normal
modes. Hence, the quasi normal frequency with smallest imaginary part
determines the late time decay of the scalar field in a black hole
geometry \cite{Konoplya:2002ky}. We find a similar decaying behavior
even in the full non-linear theory (Fig \ref{dasbasu}). 
Note that 
the scalar field falls in the black hole as expected.

A time dependent $m(t)$ changes the scalar field profile near the
boundary. As the potential vanishes near the black hole horizon,
changing the value of $m(t)$ does not have an immediate effect. Hence
a time dependent $m(t)$ creates disturbances near the boundary. The
disturbances then propagate towards horizon along a light cone.

\section{Outlook}

In this work we have taken a first step in using AdS/CFT methods
to study quantum quench across holographic critical points.  The
particular critical point we study is a mean field transition at any
finite temperature, and it is not surprising that the critical
behavior can be obtained by a Landau-Ginsburg type model of an
appropriate mode. It may appear surprising that the dynamics involves
a first order time derivative, even though the underlying theory is
Lorentz invariant. As discussed above, this is a feature of non-zero
temperature which manifests itself as a dissipative horizon in the
bulk.  However, similar techniques can be used to study other
interesting non-mean field theory transitions. In particular, the zero
temperature limit of the transition which we study is of
Berezinski-Kosterlitz-Thouless type.  Other examples of such
non-trivial transitions are discussed in \cite{bktpapers}.  The
situation is more subtle in this case : it would be worthwhile to
study quench in these systems. It would be also interesting to study
dynamics of holographic transitions associated with black branes whose
IR geometry is of Lifshitz type \cite{Goldstein:2009cv}.

As discussed above, AdS/CFT maps quantum quench in a strongly coupled
theory to a classical quench in one higher dimension. The dynamics is
completely deterministic, as opposed to e.g. Langevin dynamics - the
effect of a non-zero temperature being accounted by the presence of a
black brane in the bulk. Such systems have also been studied in the
past \cite{classquench}. In these studies, however the interesting
questions are rather different. They involve, for example, long time
behavior of fluctuations and correlation functions when averaged over
initial conditions. These questions can be addressed in our framework
as well - they would correspond to $1/N$ corrections of the dual field
theory and are left to future work. A related direction is the study
of entanglement entropy during a quench which has been studied
holographically in other contexts in
\cite{otherthermalization}-\cite{holoentanglement}. 

Finally, it would be interesting to extend our considerations to
inhomogeneous quench. In this case one interesting question relates to
defect formation as one crosses the phase transition. Some aspects of
this question has been studied numerically in \cite{saremi}. It
would be interesting to see if expected scaling properties continue to
hold, and if there is any analytical insight in terms of the
eigenmodes of the radial operator.

\appendix

\section{$\alpha=2$  quench} 

In this section we will study quantum quench in a simple
{\em relativistic} Landau-Ginsburg model. 
The relevant lagranian is \footnote{We have
  rescaled fields to set the self coupling to 1.}  
\ben {\cal
  L}=-\frac{1}{2} \dot \phi^2+ \frac{1}{2} m^2(t) \phi + \frac{1}{4}
\phi^4 + J(t) \phi
\label{22-1}
\een
In our language this is a system with $\alpha=2.$ Quench in a
holographic model without any black hole horizon, e.g. a quench in
global $AdS$, is expected to obey eqn ({22-1}) as an effective
description. This is suggested by the fact that in this latter
situation the lowest order adiabatic correction is proportional to the
second derivative of the source, as is consistent with lack of
dissipation (see e.g. the last reference in \cite{Awad}. 
This kind of model also appears in dynamics of phase transitions in
cosmology \cite{gonzales}
A
thorough discussion on those issues is left to the future.

The lagrangian (\ref{22-1}) leads to the equation of motion
\ben
\frac{d^2\phi}{dt^2}+m^2(t)\phi+\phi^3+J(t)=0
\label{22-2b}
\een
When $m^2$ and $J$ are time independent, this model has a critical point at
$m^2=J=0$. 
We study quench by seting  $m^2 = 0$ and considering a time dependent $J(t)=\tanh(vt)$.
For small $v$, breakdown of adiabaticity occurs when,
\bea
tv^{\frac{1}{4}} \sim 1
\eea
The scaling solution in the critical regime is given by,
\bea
\phi(t)= \tilde v^{\frac{1}{4}} \tilde \phi (t v^\frac{1}{4}).
\eea
As discussed in the main text this scaling relation fixes $\delta
\phi(0)$ and zero crossing time $t_{*}$.

If we continue the quench we may ask how much excess energy (or the
equivalently the amplitude of oscillation) remains in the system at 
late times. If the adiabatic approximation were valid, the
excess energy would be suppressed exponentially. However, here the
system goes through a critical point where adiabaticity is broken and
at late times the energy turns out to scale as a power of $v$.

\section{$k$ dependence of coefficients : A toy model :}

The purpose of the next two appendices is to justify the perturbation expansion
in $v^{2/5}$ used to derive the critical region scaling in section
4.2. The idea is to examine the behavior of the eigenfunctions for
small values of $k$. 
Before tackling the problem at hand we discuss a toy model in
this appendix. The next appendix discusses the eigenfunctions of the
real problem.

The toy model involves the eigenvalue equation
\ben
-\partial_\rho^2 \psi_k + V (\rho) \psi_k = k^2 \psi_k
\label{a1-1}
\een
where the potential is a step function
\begin{align}
V (\rho) &= \infty , \rho < 0 \\ 
&= -V_0 \quad, 0 < \rho<1 \\
        &= 0    \quad, \rho \geq 1
\end{align}
We want to find the eigenfunctions $\psi_k$ with continuous spectrum,
which clearly starts from zero. This potential has features which are
similar to the potential $V_0(\rho)$ in the text.

The solution for $\rho<1$ is given by
\begin{align}
 \psi_k=\frac{A(k)}{\sqrt{\pi}} \sin(\sqrt{k^2+V_0}\rho)~~~~~~~~0 < \rho < 1
\end{align}
The solution for $\rho>1$ is
\begin{align}
 \psi_k= \frac{1}{\sqrt{\pi}} \sin(k \rho+\theta(k))
\label{rhogtzero}
\end{align}
The $k$-independent coefficient in (\ref{rhogtzero}) follows from
normalization of the wave function. This may be verified by solving
the problem with a IR cutoff at some large value of $\rho = L$,
imposing Dirichlet boundary conditions there, and finally taking the
limit $L \rightarrow \infty$.

Even though there is a continuous spectrum starting at $k=0$ 
the Green's function for the
operator on the left hand side of (\ref{a1-1}) exists for generic
values of $V_0$.
However for 
\ben
\sqrt{V_0} = (n + \frac{1}{2})\pi ~~~~~~~~~n=0,1,2,\cdots
\een
there is a zero mode solution
which vanishes at $\rho = 0$ {\em and} is regular at $\rho = \infty$,
\begin{align}
  \psi_0(\rho) &=  \frac{1}{\sqrt{\pi}}\sin[(n+\frac{1}{2}) \pi \rho], \quad \rho <1 \\
 \nonumber     &= \frac{1}{\sqrt{\pi}} , \quad  \rho >1
 \end{align}
For these values of the potential, the Green's function 
diverges. The potential for $n=0$ above corresponds to a
critical point.
 
Matching two solutions at $\rho=1$ we get
\begin{align}
 A(k) \sin(\sqrt{k^2+V_0})=\sin(k +\theta(k)) \\
 A(k) \sqrt{k^2+V_0} \cos(\sqrt{k^2+V_0})=k \cos(k +\theta(k))
\end{align}
leading to 
\begin{align}
A(k) &= \frac{k}{\sqrt{k^2
    \sin^2(\sqrt{k^2+V_0})+\sqrt{k^2+V_0}\cos^2(\sqrt{k^2+V_0}})}\\ \theta(k)&=
\tan ^{-1}\left(\frac{k \tan
  \left(\sqrt{k^2+V}\right)}{\sqrt{k^2+V}}\right)-k.
\end{align}

We now examine the behavior of the wavefunctions in the small $k$
limit. 
For a generic potential $V_0\neq \frac{\pi^2}{4}$, 
the leading order behavior for small $k$ is given by
\begin{align}
A(k) &\approx k \frac{1}{\sqrt{V_0} \cos(\sqrt{V_0})}+O(k^2)
\\ \theta(k)&\approx k \left(\frac{\tan
  \left(\sqrt{V_0}\right)}{\sqrt{V_0}}-1\right)+O\left(k^3\right)
\end{align}

Now consider a function $J(\rho)$, with finite support.
The convolution of $J(\rho)$ with the
eigenfunction $\psi_k$, in the $k \rightarrow 0$ limit becomes
\begin{align}
J_k &=A(k) \int_{0}^1 dk \, \sin(\sqrt{k^2+V_0}\rho)
J(\rho)+\int_{1}^\infty dk \, \sin(k\rho+\theta(k)) J(\rho)
\\ &\approx A(k) \int_{0}^1 dk \, \sin(\sqrt{k^2+V_0}\rho)
J(\rho)+\int_{1}^\infty dk \,( k\rho+\theta(k)) J(\rho) \\ & \sim O(k)
\end{align}

Now let us discuss the critical case $V_0 = \frac{\pi^2}{4}$. 
In this case we get, upto an overall $k$-independent constant,
\begin{align}
 A(k) &\approx 1-\frac{k^2}{8}+O\left(k^4\right) \\
 \theta(k) &\approx -\frac{\pi }{2}-\frac{k}{2}+O\left(k^3\right)
\end{align}
This leads to 
\begin{align}
J_k &=A(k) \int_{0}^1 dk \, \sin(\sqrt{k^2+V_0}\rho)
J(\rho)+\int_{1}^\infty dk \, \sin(k\rho+\theta(k)) J(\rho)
\\ &\approx (1-\frac{k^2}{8}) \int_{0}^1 dk \,
\sin(\sqrt{k^2+V_0}\rho) J(\rho)+\int_{1}^\infty dk \,
(1-\frac{1}{2}(k \rho+k/2)^2) J(\rho) \\ & \sim J_0+O(k^2)
\end{align}

\section{$k$ dependence of the coefficients : the real problem} \label{kscaling}

In this appendix we will prove the scaling of quantities like $({\cal
  J}_{k}-{\cal J}_{0})$ which appear in (\ref{scalingeqn}).

The eigenvalue problem we wish to solve is,
\begin{align}
  -&\partial^2_\rho \psi_k(\rho)+V(\rho) \psi_k(\rho)=k^2 \psi_k(\rho). 
\label{appeom}
\end{align}
$k$ is a continuous parameter which may be chosen to be positive.
We will assume some simple properties for the potential $V(\rho)$ near
$\rho=\infty$, i.e., $V(\rho) \sim e^{-\rho}$ as $\rho \rightarrow \infty$.
Near the boundary $\rho = 0$ the potential $V(\rho)$ is assumed to be
sufficiently well behaved 
so that the above EOM has a solution of the form, 
\bea \psi_k(0)\sim A \rho
^{\alpha_+}+ B \rho^{\alpha_-}.  
\eea 
Where $A$ and $B$ are coefficients
of leading and subleading modes. The potential $V(\rho)$ in the
problem of interest  (i.e. the equation in a black brane background) 
satisfies these properties.  We use the boundary
condition,
\begin{align}
 A=0.
\end{align}
The normalization we use is, $\int d\rho \, \psi_k(\rho) \psi_{k'}(\rho)=\delta(k-k')$.

To solve the equation (\ref{appeom}) for small $k$, we divide the
range of $\rho$ in two overlapping parts: (i) in the ``far'' region $\rho \gg 1$
we may use $V(\rho) \approx- V_1 \exp(-\rho)$ (ii) in the ``near'' region
we can treat $k^2$ purturbatively. Then we match these
two solutions in an overlapping region to find a small $k$ solution.

In the perturbative ``far'' region:
\begin{align}
 \psi_k(\rho) = A(k)(\psi_0(\rho) + O(k^2)).
\end{align}
Here $\psi_0(\rho)$ is the solution of the $k=0$ equation with the correct
boundary conditions at $\rho = 0$ and $A(k)$ is a constant to be
determined.  Near $\rho \rightarrow \infty$, $\psi_0 \sim a
\rho+b$. The boundary condition we choose for the necessary derivative
of $\psi_0$ is such that $b=1$.

For $\rho \gg 1$ the equation (\ref{appeom}) may be approximated as,
\begin{align}
 &\partial^2_\rho \psi_k(\rho)+(V_1 \exp(-\rho)+k^2) \psi_k(\rho)=0. 
\end{align}
The real solution of the above equation is given by,
\begin{align}
 \psi_k(\rho)= B_k J_{2 i k}\left(2 \sqrt{V_1 e^{-\rho}}\right)+
 B^{*}_k J_{-2 i k}\left(2 \sqrt{V_1 e^{-\rho}}\right)
\end{align}
For $\rho \rightarrow \infty$, we have:
\begin{align}
 \psi_k \approx B_k \frac{V_1^{i k} e^{ i k \rho}}{\Gamma (2 i k+1)} +
 c.c
\end{align}
Demanding that the above solution reaches a delta function
normalizable sinusoidal solution we have,
\begin{align}
 B_k B^{*}_k=\frac{1}{4}\Gamma (2 i k+1)\Gamma (-2 i k+1).
\end{align}
This condition is needed for the normalization we use and is similar
to the choice of coefficient in (\ref{rhogtzero}) in the toy problem
considered in the previous appendix.

In the small $k$ limit we have a matching region ($1\ll \rho_0 \ll
|\log(\frac{k^2}{V1})|$) where we may expand the Bessel functions in
small argument and match with the perturbative solution. We get at
lowest order in $k$,
\begin{align}
 B_k+B^*_k=A(k) b \\
 2 i k (B_k-B^*_k)=A(k) a
\end{align}
Hence generically for small $k$, $A(k) \sim k$ ( note $b=1$). 

Things are special at the critical point, as $a=0$ there. At the
critical point $B_k=B^*_k$ and $A(k) =1 \sim o(1)$.

Now consider a function $J(\rho)$ with a compact support such that,
\begin{align}
 J_0=\int J(\rho) \psi_0(\rho) \, d \rho,
\end{align}
is finite. Now generically,
\begin{align}
 J_k=\int J(\rho) \psi_k(\rho) \, d \rho,
    \approx A(k) \int J(\rho) \psi_0(\rho) \, d \rho, \\
    \sim O(k) \, \text{as} \, k \rightarrow 0.
\end{align}
In contrast, at the critical point,
\begin{align}
 J_k&=\int J(\rho) \psi_k(\rho) \, d \rho, \\
    & \approx \int J(\rho)(A(k) \psi_0(\rho)+o(k^2)) \, d \rho, \\
    &\sim J_0+o(k^2) \text{ as } k\rightarrow 0.
\end{align}
The subleading piece comes from the perturbative $k^2$ corrections.

\section{Acknowledgements}

We would like to thank Karl Landsteiner, Satya Majumdar, Gautam Mandal, Shiraz
Minwalla, Takeshi Morita, Ganpathy Murthy, Tatsuma Nishioka, 
Omid Saremi, Alfred Shapere,
Tadashi Takayanagi, Sandip Trivedi
and especially Kristan Jensen and Krishnendu Sengupta for
disucssions. S.R.D. would like to thank Institut de Fisica Teorica at
Madrid, Tata Institute of Fundamental Research at Mumbai and Indian
Association for the Cultivation of Science at Kolkata for hospitality
during the final stages of this work.  This work is partially
supported by National Science Foundation grants PHY-0970069
and PHY-0855614.

\end{document}